\documentclass[a4paper,12pt,reqno]{amsart}

\usepackage{amsmath}
\usepackage{amsthm}
\usepackage{amssymb}
\usepackage{graphicx}
\usepackage{appendix}
\usepackage[pdftex]{hyperref}
\hypersetup{breaklinks=true,colorlinks=true,citecolor=green,urlcolor=red,linkcolor=blue,pdfpagemode=UseNone}

\theoremstyle{plain}

\newtheorem{theorem}{Theorem}[section]

\newcommand{\GL}[2]{\operatorname{GL}({#1},{#2})}

\def\COMMENT#1{}
\let\COMMENT=\footnote

\begin{document}

\title{The asymptotic complexity of matrix reduction over finite fields}
\author{Demetres Christofides}
\thanks{Demetres Christofides, School of Sciences, UCLan Cyprus, 7080 Pyla, Larnaka, Cyprus, \href{mailto:dchristofides@uclan.ac.uk}{\tt dchristofides@uclan.ac.uk}}
\thanks{This work was done during a visit to the Institut Mittag-Leffler (Djursholm, Sweden).}
\date{\today}
\begin{abstract}
Consider an invertible $n \times n$ matrix over some field. The Gauss-Jordan elimination reduces this matrix to the identity matrix using at most $n^2$ row operations and in general that many operations might be needed. 

In~\cite{AHM} the authors considered matrices in $\GL{n}{q}$, the set of $n \times n$ invertible matrices in the finite field of $q$ elements, and provided an algorithm using only row operations which performs asymptotically better than the Gauss-Jordan elimination. More specifically their `striped elimination algorithm' has asymptotic complexity $\tfrac{n^2}{\log_q{n}}$. Furthermore they proved that up to a constant factor this algorithm is best possible as almost all matrices in $\GL{n}{q}$ need asymptotically at least $\tfrac{n^2}{2\log_q{n}}$ operations.

In this short note we show that the `striped elimination algorithm' is asymptotically optimal by proving that almost all matrices in $\GL{n}{q}$ need asymptotically at least $\tfrac{n^2}{\log_q{n}}$ operations.
\end{abstract}

\maketitle

\section{Introduction}

Let $A$ be an $n \times n$ matrix with entries in some field. Our aim is to compute the inverse of $A$. The well-known Gaussian elimination does this in $O(n^3)$ steps. There are even faster algorithms than this. For example, Strassen's~\cite{Strassen} fast matrix multiplication computes the product of two $n \times n$ matrices in $O(n^{\log_2{7}})$ steps and this can be used (see e.g.~\cite{BH}) to compute the inverse of a matrix in $O(n^{\log_2{7}})$ steps as well.

In~\cite{AHM} the authors considered the complexity of matrix inversion from a different point of view. More specifically they considered methods based only on row operations and measured the complexity of matrix inversion based on the number of such operations needed. The rationale for this approach was that row operations can be implemented on existing processors far more efficiently than straight line programs. With this approach it is also the case that the problem becomes more combinatorial in nature as it is equivalent with determining the diameter of a specific Cayley graph.

Recall that if we apply some row operations to a matrix $A$ in order to reduce it to the identity matrix, then applying the same row operations to the identity matrix produces the inverse of $A$. Since our measure of complexity here is the number of row operations performed, the problems of inverting a matrix and of reducing it to the identity matrix have the same complexity. Therefore from now on we will be thinking in terms of inverting an invertible matrix or reducing it to the identity interchangeably.

The Gauss-Jordan algorithm can reduce an invertible $n \times n$ matrix to the identity in at most $n^2$ row operations (one operation per matrix element). It is easy to see that we cannot expect to improve this in general since if we have a matrix over the reals and we take all elements of the matrix to be algebraically independent, then we really do need at least $n^2$ row operations. 

In~\cite{AHM} the authors showed that if we restrict the elements of the matrix to lie in a finite field then one can improve on the Gauss-Jordan algorithm significantly. Let $\GL{n}{q}$ denote the set of all $n \times n$ invertible matrices with entries in the field of $q$ elements. The `striped elimination algorithm' of~\cite{AHM} reduces a matrix in $\GL{n}{q}$ to the identity in asymptotically at most $\frac{n^2}{\log_q{n}}$ row operations. Furthermore, it is also shown that this algorithm is optimal in the sense that for almost every matrix in $\GL{n}{q}$ we need asymptotically at least $\frac{n^2}{2\log_q{n}}$ row operations in order to reduce it to the identity. 

Our aim in this short note is to show that the `striped elimination algorithm' is optimal in a much stronger sense: Almost every matrix in $\GL{n}{q}$ needs asymptotically at least $\frac{n^2}{\log_q{n}}$ row operations in order to be reduced to the identity. More specifically, we show that following result:

\begin{theorem}\label{main}
Let $n$ be a positive integer, $q$ a prime power, and $0 < \alpha < 1$. Then using at most
\[ \frac{n^2 - 2n\log_q{n} - n - n\log_q{2} - \frac{1}{q-1}\log_q{e} - \log_q{\tfrac{1}{a}}}{\log_q{n} + \log_q{8qe}}\] 
row operations, we cannot reduce more than an $\alpha$ proportion of all matrices in $\GL{n}{q}$ to the identity matrix.
\end{theorem}

We have decided to write the bound in Theorem~\ref{main} in an explicit rather than an asymptotic format. We have not tried to optimise the lower order terms in the statements even though some of them could clearly be improved at the expense of more calculations. In any case since the `striped elimination algorithm' of~\cite{AHM} runs in a little bit more than $n^2/\log_q{n}$ steps, we have no hope to match the second order asymptotics of the algorithm using our approach.

In Section~2 we recall some elementary linear algebra facts. These are very basic facts appearing in almost every undergraduate linear algebra module. We then show that any product of elementary matrices can be written as a product of elementary matrices in a canonical way. We will use these canonical products in Section~3 in order to prove Theorem~\label{main}.

\section{Canonical products of elementary matrices}

Let $A$ be an $n \times n$ invertible matrix with entries in some field $\mathbb{F}$. We are allowed to perform the following row operations:

\begin{enumerate}
\item For $1 \leqslant i,j \leqslant n$ with $i \neq j$, interchange the $i$-th row of $A$ with its $j$-th row.
\item For $1 \leqslant i \leqslant n$ and $\lambda \in \mathbb{F}$ with $\lambda \neq 0$, multiply all elements of the $i$-th row by $\lambda$.
\item For $1 \leqslant i,j \leqslant n$ with $i \neq j$ and $\lambda \in \mathbb{F}$ with $\lambda \neq 0$, add $\lambda$ times the $i$-th row to the $j$-th row
\end{enumerate}

The result of each row operation on a matrix $A$, is exactly the same as the multiplication of $A$ from the left by an elementary row matrix. We denote these matrices by $E_{ij}, E_i(\lambda)$ and $E_{ij}(\lambda)$ corresponding to the operations (1), (2) and (3) respectively.

We perform these operations one by one until we reduce $A$ to the identity matrix. 

A crucial fact that will enable us to improve on the lower bound of~\cite{AHM} is that even though the elementary matrices do not in general commute, in many instances they do commute pairwise. The novelty of our argument is not this trivial observation per se, but how to make a good use of it. In fact we will not make full use of the commutativity, but only of the fact that if a set of row operations affect pairwise different rows, then these operations pairwise commute. So for example, even thought $E_{ij}(\lambda)$ and $E_{ik}(\mu)$ do commute, we will not use this fact. 

The other crucial fact is that even though two elementary matrices $E,E'$ might not commute we can sometimes find another elementary matrix $E''$ such that $E'E = EE''$. We will use the following instances of this observation:
\begin{equation}
E_{ij}(\mu)E_i(\lambda) = E_i(\lambda)E_{ij}(\lambda \mu)
\end{equation}
\begin{equation}
E_{ji}(\mu)E_i(\lambda) = E_i(\lambda)E_{ji}(\mu/\lambda)
\end{equation}
\begin{equation}
E_{k \ell}(\lambda)E_{ij} = E_{ij}E_{\pi_{ij}(k)\pi_{ij}(\ell)}(\lambda)
\end{equation}
\begin{equation}
E_k(\lambda)E_{ij} = E_{ij}E_{\pi_{ij}(k)}(\lambda)
\end{equation}
where in (3) and (4), $\pi_{ij}$ is the transposition interchanging $i$ and $j$. 

All of these equalities follow trivially if we consider the effects of those elementary matrices on another matrix $B$.

Suppose now that we have a product of $k$ elementary matrices. Using the above facts and observations, we will rewrite this into a new product of at most $k$ elementary matrices as follows:

Using (3),(4) and commutativity where necessary we can move all appearances of elementary matrices of the form $E_{ij}$ to the left. 

Now we look at the product of the remaining matrices and using (1),(2) and commutativity we can move all appearances of elementary matrices of the form $E_i(\lambda)$ to the left. We can also assume by commutativity that for $i < j$, every appearance of $E_i(\lambda)$ is to the left of every appearance of $E_j(\mu)$. Furthermore for each $i$, all appearances of $E_i(\lambda)$ for $\lambda \in \mathbb{R}$ now appear consecutively in the product and we can replace them with their product which is again an elementary matrix of that form.

Now we look at the product of the remaining matrices which are all of the form $E_{ij}(\lambda)$. Given an elementary matrix of the form $E_{ij}(\lambda)$, we will call the set $\{i,j\}$ its index set. We begin by partitioning these matrices into blocks as follows: We start from the left by putting each matrix into the first block one by one for as long as their index sets are disjoint. Once we reach a matrix whose index set  meets the index set of a matrix in the first block, then we put this into the second block. We now repeat by putting matrices into the second block, then create a new block as we reach a matrix whose index set meets the index set of a matrix in the second block and so on. Observe that the matrices in each block commute and so we can if we wish permute the matrices in the same block at will without changing their product. 

We now do the following modifications: Initially we do no modification in the first block. We then look at the first (from the left) matrix of the second block, if it exists, whose index set does not meet the index set of any matrix of the first block. If no such matrix exists then we do no modification to the second block either. Otherwise we move this matrix from the second block to the first, say to the last position of the first block. By repeating this for as long as it is necessary, we will end up with the situation than every matrix of the second block meets every matrix of the first block. We now move on to the third block and in the same way move matrices back to the second block for as long as they do not meet the index sets of matrices of the second block. Each time we move a matrix onto the second block we also check to see whether its index set meets an index set of a matrix of the first block. If it does not then we move it into the first block. By repeating this procedure we will end up with the situation that we will have several blocks of matrices, such that within each block the index set of matrices are disjoint while for every matrix from the second block onwards its index set will meeet the index set of at least one matrix from the previous block. This procedure is guaranteed to finish in a finite number of steps. For example, giving to each matrix as value the number of the block in which it appears, then the sum of the values of the matrices reduces after each step of the procedure. Furthermore this number is a non-negative integer and so the procedure cannot go on forever.

We have now finished in rewriting the initial product of elementary matrices as a new product with some specific properties. We call any such product a canonical product of elementary matrices. More specifically, we say that a product $E_1E_2 \cdots E_k$ of elementary matrices $n\times n$ matrices is {\bf canonical} if there exist non-negative integers $r,r_0,r_1,\ldots,r_s$ such that 
\begin{itemize}
\item[(a)] Each $E_1,\ldots,E_r$ is equal to $E_{ij}$ for some $i,j$. 
\item[(b)] Each $E_{r+1},\ldots,E_{r+r_0}$ is equal to $E_i(\lambda)$ for some $i,\lambda$. Furthermore, if $E_t = E_i(\lambda)$ and $E_{t'} = E_{i'}(\lambda')$ where $r+1 \leqslant t < t' \leqslant r+r_0$ then $i < i'$.
\item[(c)] Each $E_k$ with $k > r+r_0$ is equal to $E_{ij}(\lambda)$ for some $i,j,\lambda$. Furthermore, if we write $I_k = \{i,j\}$ for the index set of this elementary matrix and define $r_i' = r + r_0 + r_1 + \cdots + r_i$ for each $0 \leqslant i \leqslant s$ then the following holds: 
\begin{itemize}
\item[(i)] For each $1 \leqslant i \leqslant s$, we have that the index sets $I_{r_{i-1}'+1}, \ldots, I_{r_i'}$
are pairwise disjoint.
\item[(i)] For each $2 \leqslant i \leqslant s$, and each $t \in [r_{i-1}'+1,r_i']$ there is a $t' \in [r_{i-2}'+1,r_{i-1}']$ with $I_t \cap I_{t'} \neq \emptyset$. 
\end{itemize} 
\end{itemize}

\section{Proof of Theorem~\ref{main}}

Suppose that every matrix in $\GL{n}{q}$ can be reduced to the identity matrix using at most $k$ row-operations. From our results in Section~2, it follows that every such matrix can be written as a canonical product of at most $k$ elementary matrices. 

This canonical product starts with a product of matrices of the form $E_{ij}$. Their product is a permutation matrix, so there are at most 
\[ n! \leqslant n^n = q^{n \log_q{n}}\] 
different product that can be obtained so far.

The canonical product continues with a product of matrices of the form $E_i(\lambda)$. There are $2^n$ ways to choose which indices $i$ appear in the matrices of this product. For each such matrix, there is a total of $q-1$ ways to choose $\lambda$. So in total the product of those matrices can be formed in at most
\[ 2^n (q-1)^n \leqslant q^{n \log_q{2} + n}\]
ways.

Finally, there are at most $k$ more matrices to consider, all of the form $E_{ij}(\lambda)$. These matrices will appear into blocks of $r_1,r_2,\ldots,r_s$ matrices for some non-negative integer $s$ and some positive integers $r_1,\ldots,r_s$ with $r_1 + \cdots + r_s \leqslant k$. Within each block the index sets of the matrices used are all disjoint, while the index set of every matrix of a block meets the index set of at least one matrix of the previous block. 

There are exactly $2^k$ ways in order to choose the numbers $r_1,\ldots,r_s$. To see this observe that given positive integers $r_1,\ldots,r_s$ such that $r_1 + \cdots + r_s \leqslant k$, these determine the subset $\{r_1,r_1+r_2,\ldots,r_1+\cdots + r_s\}$ of $\{1,2,\ldots,k\}$. Conversely, given any subset $\{x_1,\ldots,x_s\}$ of $\{1,2,\ldots,k\}$ with $x_1 < x_2 < \cdots < x_s$ then $r_1=x_1,r_2 = x_2-x_1,\ldots,r_s = x_s - x_{s-1}$ are positive integers with $r_1 + \cdots + r_s \leqslant k$. Furthermore these two maps between tuples of positive integers summing up to at most $k$, and subsets of $\{1,2,\ldots,k\}$ are inverses of each other and so indeed the number of ways to choose $r_1,\ldots,r_s$ is equal to $2^k = q^{k\log_q{2}}$. 

Suppose now that $r_1,\ldots,r_s$ have been chosen. There are at most $q^{r_1}n^{2r_1}$ ways to choose the first $r_1$ matrices. Here the $q^{r_1}$ is for the choose of $\lambda$'s and the $n^{2r_1}$ for the choice of indices. Having chosen those, there are at most $q^{r_2}(4r_1n)^{r_2}$ ways to choose the second $r$ matrices. This is because when choosing each of the $r_2$ matrices of this block, there are $2$ ways to choose which element of its index set will meet the index set of a matrix from the previous block, there are at most $2r_1$ ways to choose that element, and there are at most $n$ ways to choose the other element. Similarly, there are $q^{r_3}(4r_2n)^{r_3}$ ways to choose the matrices of the third block and so on. 

So in total for fixed $r_1,r_2,\ldots,r_s$ there are

\begin{align*} 
q^{r_1}n^{2r_1}(4qr_1n)^{r_2} \cdots (4qr_{s-1}n)^{r_s} &\leqslant n^{2r_1 + r_2 \cdots + r_s} (4q)^{r_1 + \cdots + r_s}r_1^{r_2} \cdots r_{s-1}^{r_s}r_s^{r_1} \\
&\leqslant n^{n+k}(4q)^k r_1^{r_2} \cdots r_{s-1}^{r_s}r_s^{r_1} \\
&= q^{n\log_q{n} + k\log_q{n} + 2k\log_q{2} + k}r_1^{r_2} \cdots r_{s-1}^{r_s}r_s^{r_1}.
\end{align*}

ways to form this product.

However many of those products give rise to the same matrix. In particular, the order in which we pick the matrices of the first block does not matter as it will give up the same product. The same holds for the order of the matrices within each block. So for each $r_1,\ldots,r_s$, each possible product has been appeared in the above calculation at least $r_1! \cdots r_s!$ times.

We now observe that

\[ r_1! \cdots r_s! \geqslant \left( \tfrac{r_1}{e}\right)^{r_1} \cdots  \left( \tfrac{r_s}{e}\right)^{r_s}\geqslant \frac{r_1^{r_1} \cdots r_s^{r_s}}{e^k}.\]

Since the function $x \mapsto \log_q{x}$ is an increasing function of $x$, the rearrangement inequality shows that 

\[ r_1 \log_q{r_1} \cdots r_s \log_q{r_s} \geqslant r_2\log_q{r_1} + \cdots + r_s\log_q{r_{s-1}} + r_1\log_q{r_s}\]

and so 

\[ r_1^{r_1} \cdots r_s^{r_s} \geqslant r_1^{r_2} \cdots r_{s-1}^{r_s}r_s^{r_1}.\]

So in total, for each $r_1,\ldots,r_s$ there are at most 

\[ q^{n\log_q{n} + k\log_q{n} + 2k\log_q{2} + k + k\log_q{e}}\]

different products that can be formed.

%
%
%
%
%
%
%
%

So putting everything together, there is a total of at most

\begin{equation}\label{eqn} 
q^{(k+2n)\log_q{n} + (3k+n)\log_q{2} + n + k + k\log_q{e}}
\end{equation}

distinct matrices that can arise from canonical products of at most $k$ matrices from $\GL{n}{q}$.

Finally, it is not difficult to see that 

\[ |\GL{n}{q}| = \prod_{k=0}^{n-1} (q^n - q^k) = q^{n^2} \prod_{k=0}^{n-1} (1 - q^{k-n}) = q^{n^2}\prod_{r=1}^n\left(1 - \frac{1}{q^r} \right).\]

But

\[ \prod_{r=1}^n\left(1 - \frac{1}{q^r} \right)= e^{\sum_{r=1}^n \log(1 - q^{-r})} \geqslant e^{-\sum_{r=1}^n q^{-r}} \geqslant e^{-\frac{1}{q-1}}\]

and so there are at least 

\[e^{-\frac{1}{q-1}} q^{n^2} = q^{n^2 - \tfrac{1}{q-1}\log_q{e}}\] 

invertible $n \times n$ matrices with entries in $\mathbb{F}_q$. This, together with~\eqref{eqn}, complete the proof of Theorem~\ref{main}.


\begin{thebibliography}{9}
\bibitem{AHM} D. Andr\'en, L. Hellstr\"om\ and\ K. Markstr\"om, On the complexity of matrix reduction over finite fields, Adv. in Appl. Math. {\bf 39} (2007), 428--452.

\bibitem{BH} J. R. Bunch\ and\ J. E. Hopcroft, Triangular factorization and inversion by fast matrix multiplication, Math. Comp. {\bf 28} (1974), 231--236.

\bibitem{Strassen} V. Strassen, Gaussian elimination is not optimal, Numer. Math. {\bf 13} (1969), 354--356.
\end{thebibliography}
\end{document}